\documentstyle[multicol,aps,prb,floats,eqsecnum]{revtex}

\newcommand{\calD}{{\cal D}}

\newcommand{\tr}{{\rm tr}}
\newcommand{\bfn}{{\bf n }}
\newcommand{\bfN}{{\bf N }}
\newcommand{\psibar}{{\overline{\psi}}}

\newcommand{\zbar}{{\bar z}}
\newcommand{\partialbar}{{\bar\partial}}
\newcommand{\intspace}{\!\!\!\!\!}

\begin{document}
\draft                                        
\preprint{}                                   
\title{ Strong disorder effects of a Dirac fermion with
a random vector field } 
\author{ Takahiro Fukui\cite{Email}}
 \address{Theoretical Physics, University of Oxford, 
          1 Keble Road, Oxford OX1 3NP, UK}
\date{\today}
\maketitle
\begin{abstract}
We study a Dirac fermion model with a random vector field,
especially paying attention to a strong disorder regime.
Applying the bosonization techniques, 
we first derive an equivalent sine-Gordon model, 
and next average over the random vector field using the replica trick.
The operator product expansion based on the replica action leads to 
scaling equations of the coupling constants (``fugacities'')
with nonlinear terms,
if we take into account the fusion of the vertex operators.
These equations are converted into a nonlinear diffusion equation
known as the KPP equation. By the use of
the asymptotic solution of the equation, we calculate 
the density of state, the generalized inverse participation ratios,
and their spatial correlations.
We show that results known so far are all derived
in a unified way from the point of view of the renormalization group.
Remarkably, it turns out that the scaling exponent obtained in this paper
reproduces the recent numerical calculations of the density 
correlation function.
This implies that the freezing transition has actually 
revealed itself in such calculations.
\end{abstract}


\begin{multicols}{2}

\section{Introduction} \label{s:Int}

A Dirac fermion model has proved to be quite useful,
even in condensed matter physics, as an effective 
theory for various kinds of phenomena,
e.g, the integer quantum Hall transition,\cite{LFSG,MCW,Ber} 
$d$-wave superconductors,\cite{Lee,NTW,SFBN,BCSZ,ASZ} etc.
It is also involved with statistical mechanical models
such as the Ising model\cite{Kog} or the XY-model.\cite{MudWen}
Especially, including randomness, this model has clarified 
remarkable aspects of disordered systems, for example,
$\ln\ln\tau$ behavior of the specific heat of the random-bond
Ising model \cite{DotDot}
and multifractal scaling dimensions.\cite{LFSG,MCW,CMW,CCFGM}

Many problems, however, still remain to be explored.
Especially, the critical theory of a Dirac fermion 
with generic disorder, 
which is believed to be an effective theory of 
the integer quantum Hall transition, is still missing.\cite{LFSG}
In this case, coupling constants describing disorder strength flow to
a strong coupling regime, which we cannot reach 
by weak coupling approaches.
Even simpler model with a random vector field only,
whose zero energy state can be obtained exactly for 
any realization of randomness, yields non-trivial 
strong disorder phase, 
recognized through the calculation of the generalized
inverse participation ratios (IPR).\cite{CMW,CCFGM}

On the other hand, 
Carpentier and Le Doussal have recently  developed a theory for 
the XY-model with a random gauge field.\cite{CarDouN} 
It has been shown that
a naive replica theory for this model concludes a reentrant 
transition,\cite{CarOst,RSN}
which is not consistent to the numerical calculations\cite{XYCal}
and to exact arguments by the use of the Nishimori-line.\cite{OzeNis}
Although this inconsistency has been a long standing problem,
underlying physics has recently been clarified.\cite{XYDev,MudWen}
In particular, Scheidl\cite{Sch} has pointed out the importance of a role
which vortices with higher charges play.
Carpentier and Le Doussal have derived new scaling equations 
for the Coulomb gas model taking 
into account  ``fusion'' of vortices, from which 
higher charge vortices are generated consistently.
The renormalization group (RG) equations they have derived have
intimate relationship with the freezing transition of spin glasses
(the random energy model).\cite{Der}

By the use of the fact that the exact zero energy wave function
of a Dirac fermion model with a random vector field is 
equivalent to the vertex operator 
of the boson field theory,\cite{CarDouE}
Carpentier and Le Doussal have applied their RG
method to a Dirac fermion model, calculating the IPR.
Variational method has also been applied
in order to calculate the density of state (DOS).\cite{HorDou}
It has turned out that these quantities shows peculiar
behavior in a strong coupling regime 
due to the freezing transition.

Motivated by these developments, we rederive, in this paper, 
scaling equations for a replicated sine-Gordon model 
with a random vector field,
which is the bosonized model of a Dirac fermion with a random
vector field. The main point is that we take into account 
the fusion of the vertex operators.
In order to derive scaling equations at one-loop order, we use 
the operator product expansion (OPE) techniques.
Based on these, we calculate the DOS, the
IPR, and their spatial correlations of a Dirac fermion model.
We show that it is possible to derive all results known so far
in a unified way from the RG point of view. 
Moreover, it turns out that the scaling exponent obtained in this paper
reproduces the recent numerical calculations by Ryu and Hatsugai
of the density correlation function.\cite{RyuHat}
This implies that the freezing transition has manifestly revealed itself 
in their calculations.

This paper is organized as follows:
In the next section, a Dirac fermion model is introduced 
and a corresponding sine-Gordon model is derived.
In Sec. \ref{s:Sca}, 
the OPEs are calculated, which lead to scaling equations, 
by taking account of the fusion of the vertex operators, and 
in Sec. \ref{s:KPP}, they are converted into more convenient form,
known as the KPP equation.\cite{KPP,Bra,EbeSaa}
By the use of the asymptotic solution of the equation,
the DOS and the IPR are calculated in Sec. \ref{s:DOS}.
A summary and discussions are given in Sec. \ref{s:Sum}.

\section{Dirac fermion with random vector field}\label{s:Dir}

In this section, we first introduce a Dirac fermion model including
a random vector field and give formulae to calculate 
the DOS, etc. Next, using the bosonization techniques, we convert
the Dirac action into an equivalent sine-Gordon action.
The latter is the model we directly investigate in this paper.

\subsection{Dirac fermion model}
The action functional is defined by
\begin{eqnarray}
S=\int d^2x
\left[
\psibar i\gamma_\mu\left(\partial_\mu-iA_\mu\right)\psi
-iy\psibar\psi
\right] ,
\label{DirAct}
\end{eqnarray}
where $y=\omega-iE$, $\gamma_\mu=\sigma_\mu$ (the Pauli matrices), 
and $A_\mu$ denotes a quenched random vector field
with the following gaussian probability distribution
\begin{eqnarray}
P[A_\mu]\propto \exp\left(-\frac{1}{2\pi g}\int d^2xA_\mu^2\right) .
\end{eqnarray}
Here we include an extra factor $\pi$ into the definition of
the distribution width $g$ for later convenience.
The Green function is computed as 
\begin{eqnarray}
{\rm Im}~\tr\langle&&\psi(x)\psibar(x)\rangle
\nonumber\\&& 
=\tr
\langle x|
\frac{\omega}{[i\gamma_\mu(\partial_\mu-iA_\mu)-E]^2+\omega^2}
|x\rangle ,
\end{eqnarray}
where tr is the trace with respect to the two-component spinor.
Let us further define the $q$th power of the Green function
\begin{eqnarray}
\Gamma^{(q)}(x)=
\left({\rm Im}~\tr\langle\psi(x)\psibar(x)\rangle\right)^q  .
\end{eqnarray}
Then, the following relation is valid
\begin{eqnarray}
&&\omega^{q-1}\Gamma^{(q)}(x)
\nonumber\\&&\quad 
\rightarrow C_q\sum_n|\Psi_n(x)|^{2q}\delta(E_n-E),
\quad (\omega\rightarrow+0) , 
\label{GreFunFac}
\end{eqnarray}
where $\Psi_n(x)$ is the eigenstate of the
Hamiltonian with energy $E=E_n$, 
namely, $i\gamma_\mu(\partial_\mu-iA_\mu)\Psi_n=E_n\Psi_n$,
and $C_q$ is a numerical constant $C_q=\pi(2q-3)!!/(2q-2)!!$.
Field-theoretically, the ensemble-averaged DOS $\rho(E)$, 
the generalized IPR $P^{(q)}(E)$,
and their spatial correlations $Q^{(q_1,q_2)}(x,y,E)$
are defined by\cite{MCW,Weg}, in the limit $\omega\rightarrow+0$,
\begin{eqnarray}
&&\rho(E)=\overline{\Gamma^{(1)}(x)}/C_1,
\nonumber\\
&&P^{(q)}(E)
=\omega^{q-1}\overline{\Gamma^{(q)}(x)}/\left[C_q\rho(E)\right],
\nonumber\\
&&Q^{(q_1,q_2)}(x-y,E)
\nonumber\\&&\quad 
=\omega^{q_1+q_2-1}\overline{\Gamma^{(q_1)}(x)\Gamma^{(q_2)}(y)}/
\left[C_{q_1}C_{q_2}\rho(E)\right] .
\label{DefDosEtc}
\end{eqnarray}
Here we have assumed that the ensemble-average recovers the 
translational invariance and hence
$\overline{\Gamma^{(q)}(x)}$
does not depend on $x$ and 
$\overline{\Gamma^{(q_1)}(x)\Gamma^{(q_2)}(y)}$ depends on 
the separation $x-y$ only.

\subsection{Sine-Gordon model}

Bosonization of (\ref{DirAct}) yields the following 
sine-Gordon action
\begin{eqnarray}
S=\int\frac{d^2x}{4\pi}
\left[
\frac{1}{2}(\partial_\mu\phi)^2
+2iA_\mu\epsilon_{\mu\nu}\partial_\nu\phi
-y\cos\phi
\right] .
\end{eqnarray}
By the use of the replica trick, ensemble-average 
over the vector field $A_\mu$ gives
\begin{eqnarray}
\overline{Z^m}=\int\calD\phi e^{-S^{(m)}},
\end{eqnarray}
with 
\begin{eqnarray}
&&
S^{(m)}
\nonumber\\&& 
=\int\frac{d^2x}{4\pi}
\left[
\frac{1}{2}\sum_{a,b=1}^m
\partial_\mu\phi_a G_{ab}\partial_\mu\phi_b
-y\sum_{a=1}^m\cos\phi_a
\right] ,
\label{NaiSinGor}
\end{eqnarray}
where $a$, $b$ denote $m$ replicas and
\begin{eqnarray}
G_{ab}=\frac{1}{K}\delta_{ab}+g .
\end{eqnarray}
The bare coupling constant is given by $K=1$,
since the present system is a free fermion model.

So far we have derived a
replicated sine-Gordon model (\ref{NaiSinGor})
by the bosonization of the Dirac fermion model.
We also reach the same model from the XY-model with a random gauge field
as a result of applying the Villain approximation, 
``integrating" over vortices, 
and performing the quenched average with a replica trick.\cite{CarDouN}
In this case, the coupling constant $K$ is $K=J/T$,
where $J$ and $T$ denotes, respectively, the spin coupling constant
and the temperature.
The problem had been that the RG analysis
based on this action (\ref{NaiSinGor})
predicts a reentrant transition, which 
contradicts numerical studies\cite{XYCal} 
and exact arguments by the use of
the Nishimori-line.\cite{OzeNis}

Recent developments,\cite{XYDev,Sch,CarDouN}
however, have made it possible to describe the random XY
model within a framework of the replica method, 
without resort to the replica symmetry breaking. 
In the language of the sine-Gordon model, 
the main point is that higher charge vertex operators,
which are not included in the above action but are generated 
in the process of the renormalization, are taken into account.
The next section is devoted to the reformulation of the method.

\section{Scaling equations}\label{s:Sca}

Carpentier and Le Doussal have recently proposed improved 
scaling equations for the Coulomb gas model.\cite{CarDouN}
They have taken into account the fusion of vortices, 
which results in scaling equations with nonlinear terms.
Without these terms, the flow of the coupling constant $y$
(also  called ``fugacity'' in the context of the Coulomb gas model) 
is governed by the
exponent $2-x$, where $x$ is the dimension of $\sum_a\cos\phi_a$
in Eq. (\ref{NaiSinGor}).
The nonlinearity of the scaling equations, however, 
has turned out to play a crucial role in the RG flow.
In what follows, we rederive the scaling equations for 
the present sine-Gordon model by the use of the OPE techniques.

We start with the same sine-Gordon action but 
with generalized fugacities,
\begin{eqnarray}
S=\int\frac{d^2x}{4\pi}
\left[
\frac{1}{2}\partial_\mu\phi_a G_{ab}\partial_\mu\phi_b
-\sum_{\bfn\ne0} Y(\bfn)e^{in_a\phi_a}
\right] ,
\label{Act}
\end{eqnarray}
where $a,b=1,2,\cdots m$, $\bfn^t=(n_1,n_2,\cdots,n_m)$ 
is a $m$ component vector with integer elements $n_a$
which will be specified momentarily.
The vector \bfn denotes a ``charge'' of the vertex operator,
so that let us define a total charge $n=\sum_an_a$.
Initially, the bare fugacity is 
$Y(\bfn)=y_1(\delta_{\bfn,\bfn_1}+\delta_{\bfn,-\bfn_1})$, 
where the vector $\bfn_1$ is such that
only one of its elements is $1$ and others are zero.
In this case,
$\sum_{\bfn\ne0} Y(\bfn)e^{in_a\phi_a}
=2y_1\sum_{a=1}^m\cos(\phi_a) $,
and the action (\ref{Act}) 
reduces to the normal replicated sine-Gordon
action (\ref{NaiSinGor}).

The unperturbed correlation function reads
\begin{eqnarray}
\langle\phi_a(x)\phi_b(y)\rangle
&&=\left(\frac{G}{4\pi}\right)^{-1}_{ab}(-\partial^2)^{-1}(x-y)
\nonumber\\
&&=G^{-1}_{ab}\ln|x-y|^{-2} .
\label{TwoPoiFun}
\end{eqnarray}
The inverse of the matrix $G$ is easy to calculate,
\begin{eqnarray}
G^{-1}_{ab}&&=K\delta_{ab}-\frac{g K^2}{1+mg K}
\nonumber\\
&&\rightarrow K\delta_{ab}-g K^2,
\end{eqnarray}
where we have taken the replica limit $m\rightarrow0$
in the last equation.
By using the correlation function (\ref{TwoPoiFun}), we readily find,
\begin{eqnarray}
\langle e^{in_a\phi_a(z)}e^{-in_a\phi_a(0)} \rangle
=\frac{1}{|z|^{2x(\bfn)}} ,
\end{eqnarray}
where the dimension $x(\bfn)$ of $e^{in_a\phi_a}$ is
\begin{eqnarray}
x(\bfn)
&&
=\bfn^tG^{-1}\bfn
\nonumber\\
&&= K|\bfn|^2 -g K^2n^2 ,
\end{eqnarray}
with $|\bfn|^2=\bfn\cdot\bfn$ and $\bfn\cdot\bfn'=\sum_an_an_a'$.

The OPE of the vertex operators 
in the case $\bfn+\bfn'=\bfn''\ne0$ is given by
\begin{eqnarray}
e^{in_a\phi_a(z)}e^{in'_a\phi_a(0)}
=\frac{1}{|z|^{x(\bfn,\bfn')} }
e^{in''_a\phi_a(0) }  ,
\label{OPEI}
\end{eqnarray}
with
\begin{eqnarray}
x(\bfn,\bfn')&&=x(\bfn)+x(\bfn')-x(\bfn'')
\nonumber\\
&&
=-2\bfn^{t}G^{-1}\bfn'
\nonumber\\
&&
=2\left(g K^2nn'-K\bfn\cdot\bfn'\right) , 
\end{eqnarray}
where we have used the fact that the matrix $G$ is symmetric.
On the other hand, 
in the case $\bfn+\bfn'=0$, we find
\begin{eqnarray}
e^{in_a\phi_a(z)}e^{-in_a\phi_a(0)}
&&\sim\frac{1}{|z|^{2x(\bfn)}}
e^{in_a(z\partial+\zbar\partialbar)\phi_a(z)} 
\nonumber\\
&&\sim\frac{-1}{|z|^{2x(\bfn)-2}}
\partial\phi_a (\bfn\bfn^t)_{ab} \partialbar\phi_b ,
\label{OPEII}
\end{eqnarray}
where $z,\zbar=x_1\pm ix_2$, 
$\partial=\partial_z$ and $\partialbar=\partial_\zbar$.

First, the OPE (\ref{OPEI})
leads us directly to the scaling equations,
\begin{eqnarray}
&&
\frac{dY(\bfn)}{dl}=\left[2-x(\bfn)\right]Y(\bfn)+\frac{1}{4}
\sum_{{\bfn'+\bfn''=\bfn}\atop{\bfn'\ne0, \bfn''\ne0}}
Y(\bfn')Y(\bfn''),
\nonumber\\&& 
\label{ScaEqu1}
\end{eqnarray}
where $l=\ln L$ with the system size $L$.

So far we have not specified the vector charges $\bf n$.
The initial condition at $l=0$ is such that 
there exist $\bfn_1$-type vectors only. 
However, the ``fusion" of the vertex operators (\ref{OPEI}) yields
higher charge vectors in the process of the renormalization.
Let us fix the charge $n$, and compute the 
scaling dimensions of the vertex operators. 
Consider the case with $n=1$ for simplicity. Then, we have
$\bfn^t=(1,0,0,\cdots)$, 
$(1,1,-1,0\cdots)$, and $(2,-1,0,\cdots)$,
for example, whose dimensions are, respectively,
$K-gK^2$, $3K-gK^2$, and $5K-gK^2$.
It is readily seen that the vectors of the first type 
have the most relevant dimension.
In a similar way, given a charge $n$, we find that 
the most relevant vectors are
the ones that have $|n|$ 1's $(-1$'s) for a positive (negative) $n$.

In what follows, we restrict ourselves to these most relevant ones,
$\bfn$ with $n_a=0,1$ and with $n_a=0,-1$, and therefore,
vectors whose charges lie between $\pm m$ are taken into account.
Let us define a set of vectors 
$\bfN_q$ which includes vectors 
$\bfn$ with $n_a=0,1$ and $\sum_an_a=q$.
The scaling dimensions depend, in this case, only on $|n|$,
so that let us define, for $n>0$
\begin{eqnarray}
x(\bfn)= x_n\equiv Kn-g K^2n^2 .
\label{RisVerDim}
\end{eqnarray}
Then the scaling equations (\ref{ScaEqu1}) decouple
into positive and negative charge sectors, which obey the same
scaling equations with the same initial condition. 
Thus, it turns out that  $Y(-\bfn)=Y(\bfn)$.
Moreover, $Y(\bfn)$ depends on $n$ only. 
Denoting $Y(\bfn)=Y(-\bfn)=y_n$ and counting the multiplicity
in the summation over $\bfn'$ and $\bfn''$ in Eq. (\ref{ScaEqu1}), 
we end up with a closed set of scaling equations
\begin{eqnarray}
&&
\frac{dy_n}{dl}=\left(2-Kn+g K^2n^2\right)y_n+\frac{1}{4}
\sum_{n'=1}^{n-1}{n\choose n'}y_{n'}y_{n-n'}.
\nonumber\\&& 
\label{RGEquFug}
\end{eqnarray}

Next consider the OPE (\ref{OPEII}). 
The exponent in this equation is
$2x_n-2=2(n-gn^2-1)$,
where we have used the free fermion condition $K=1$.
It should be noted that in a weak disorder regime,
a naively replicated model including $y_1$ only could describe the 
critical behavior correctly. The exponent of this 
fusion is $2x_1-2=-2g$, which is definitely
{\it negative}, and therefore,
gives rise to no ultraviolet singularity.
Hence, the kinetic term should not be renormalized.
This may be valid in a strong coupling regime,
since large $g$ tends to make the exponent negative,
although in some range of $n$, positive exponents appear
if one includes higher charge vectors.
Taking these into account, we postulate that the kinetic term
is not renormalized in any regime of interest
$K \stackrel{\textstyle<}{\sim}1$, in what follows.

The Green function can be described by the replicated bose fields as
$\Gamma^{(q)}\sim
\langle\cos\phi_1\rangle\langle\cos\phi_2\rangle\cdots
\langle\cos\phi_q\rangle$.
If one keeps the most relevant field, one finds
\begin{eqnarray}
&&
\overline{\Gamma^{(q)}(x)}\sim
\langle\cos\left(n_a\phi_a(x)\right)\rangle, 
\nonumber\\
&&
\overline{\Gamma^{(q_1)}(x)\Gamma^{(q_2)}(y)}
\nonumber\\&&\quad 
\sim 
\langle\cos\left(n'_a\phi_a(x)\right)
\cos\left(n''_b\phi_b(y)\right)\rangle, 
\label{GreBos}
\end{eqnarray}
after the ensemble average, where
$\bfn\in\bfN_q$, $\bfn'\in\bfN_{q_1}$, and $\bfn''\in\bfN_{q_2}$.

\section{The KPP equation}\label{s:KPP}

So far we have derived the scaling equations
with nonlinear terms due to the fusion of the vertex operators.
Remarkably, it has been shown\cite{CarDouN} that the nonlinear terms
are involved with the freezing transition 
of the random energy model.\cite{Der}
This section is devoted to the review of
converting the scaling equations (\ref{RGEquFug}) 
into the KPP equation,\cite{KPP,Bra,EbeSaa}
according to Carpentier and Le Doussal.\cite{CarDouN}
First, make the Mellin transformation
$y_n(l)=\int_0^\infty dzz^{n-1}P(z,l)$,
which is converted, by the use of the variable $z=e^u$,
into more convenient form for the 
present purpose as
\begin{eqnarray}
y_n(l)=\int_{-\infty}^\infty\intspace due^{nu}P(u,l) .
\label{MelTra}
\end{eqnarray}
In the $u$-space, $n$ can be replaced by 
$n\rightarrow-\partial_u$, and then we have
\begin{eqnarray}
\partial_lP(u,l)
=&&\left[2+K\partial_u+g K^2\partial_u^2\right]P(u,l)
-\frac{1}{2}y_0P(u,l)
\nonumber\\&& 
+\frac{1}{4}
\int_{-\infty}^\infty\intspace du_1
\int_{-\infty}^\infty\intspace du_2
\delta\left(u-\ln(e^{u_1}+e^{u_2})\right)
\nonumber\\
&&\times
P(u_1,l)P(u_2,l).
\end{eqnarray}
Integration over $u$ of the both sides of the equation yields
\begin{eqnarray}
\partial_ly_0=2y_0-\frac{1}{4}y_0^2 .
\end{eqnarray}
It turns out that $y_0$, which is just the normalization 
of the function $P(u,l)$, flows as $y_0\rightarrow8$.
Therefore, setting $y_0=8$ gives
\begin{eqnarray}
\partial_lP(u,l)
=&&\left[K\partial_u+g K^2\partial_u^2\right]P(u,l)
-2P(u,l)
\nonumber\\&& 
+\frac{1}{4}
\int_{-\infty}^\infty\intspace du_1
\int_{-\infty}^\infty\intspace du_2
\delta\left(u-\ln(e^{u_1}+e^{u_2})\right)
\nonumber\\
&&\times
P(u_1,l)P(u_2,l).
\label{ScaEquP}
\end{eqnarray}

Although it may be difficult to solve this equation analytically,
a crucial observation was made by 
Carpentier and Le Doussal as follows:\cite{CarDouN} 
Define a generating functional of $y_n$ by
\begin{eqnarray}
G(x,l)&&=1-
\left\langle\exp\left(-ze^{-x+Kl}\right)\right\rangle_P
\nonumber\\
&&=\frac{1}{8}\sum_{n=1}^\infty\frac{(-)^{n+1}}{n!}y_n(l)
e^{-n(x-Kl)} ,
\label{DefG}
\end{eqnarray}
where $z=e^u$, and we have introduced the notation
$\langle A\rangle_P\equiv\int_{-\infty}^\infty \frac{du}{8}P(u,l)A$,
since the function $P(u,l)$ can be interpreted, 
after normalized to unity,
as the probability distribution of the fugacity.\cite{CarDouN}
What is remarkable is that the function $G(x,l)$ obeys the 
following equation
\begin{eqnarray}
\frac{1}{2}\partial_lG=
D\partial_x^2G+G(1-G) ,
\label{FKPP}
\end{eqnarray}
where $D=gK^2/2$.
This equation is known as the KPP equation.\cite{KPP,Bra,EbeSaa}
Since $y_0=8$, the boundary condition is 
$G(-\infty,l)=1$ and $G(\infty,l)=0$.

\section{The DOS and the IPR}\label{s:DOS}

In order to calculate the dominant exponents of the DOS, etc in 
Eq. (\ref{DefDosEtc}), we need to know the scaling exponents
of the local composite operators $\Gamma^{(q)}$.
Basically they are given by (\ref{RisVerDim}) 
through the relation (\ref{GreBos}). 
However, it will turn out that the nonlinear terms
of the scaling equations 
induce nontrivial dynamical scaling exponents in a strong
disorder regime.
Therefore, we first have to read the dynamical scaling exponents 
from the solution of the KPP equation, 
and next calculate various exponents of the DOS, etc.

\subsection{Dynamical scaling exponents of the vertex operators}

First of all, let us summarize the remarkable selection rule
of the front velocity of the KPP equation (\ref{FKPP}).
Provided that $G(x,l)$ has a boundary condition 
$G(-\infty,l)=1$ and $G(\infty,l)=0$ as well as an initial condition
$G(x,l=0)\sim e^{-\mu x}$ at large $x$. 
In this case, the KPP equation admits the following 
traveling solution\cite{Bra}
\begin{eqnarray}
G(x,l)=g(x-m(l)), \quad\mbox{for}\quad l\rightarrow \infty .
\end{eqnarray}
The solution tells that the wavefront (a kink) travels with a speed 
$\partial_lm(l)$ if we interpret $l$ as time.
Neglecting the nonlinear term $G^2$ and assuming the 
form of the solution as $\exp\left[-\mu(x-vl)\right]$,
we expect that the velocity $v$, defined by $m(l)=vl$,
is given by $v=2(D\mu+1/\mu)$.
Actually, this is valid, as long as $\mu$ is small enough.
However, it has been shown\cite{Bra} that if $\mu$ is larger than
$\mu_{cr}=1/\sqrt{D}$, the relation does not hold any longer,
and the velocity is a constant (the minimum value) independent of $\mu$.
To be precise, $m(l)$ is given by\cite{Bra}
\begin{eqnarray}
&&
m(l)=\left\{
\begin{array}{ll}
2(D\mu+1/\mu)l+O(1)
&
\mbox{for}\quad \mu<1/\sqrt{D},
\\
\sqrt{D}\left(4l-\frac{1}{2}\ln l \right)+O(1) 
&
\mbox{for}\quad \mu=1/\sqrt{D} ,
\\
\sqrt{D}\left(4l-\frac{3}{2}\ln l \right) +O(1)
&
\mbox{for}\quad  \mu>1/\sqrt{D} .
\end{array}
\right.
\nonumber\\
&&
\label{FroVel}
\end{eqnarray}

Next task is to obtain the 
behavior of the fugacities $y_n$, which are related to the above 
solution of $G(x,l)$ via Eqs. (\ref{MelTra}) and (\ref{DefG}). 
To this end, notice that 
the front velocity above is related to a typical value of
$u$ via the relation,\cite{CarDouN,DerSpo}
\begin{eqnarray}
u_{typ}&&\equiv[\ln z]_P
\nonumber\\
&&
=\int_{-\infty}^\infty\intspace dx
\left[
\exp\left(-e^{-x+Kl}\right)-
\left\langle\exp\left(-ze^{-x+Kl}\right)\right\rangle_P
\right]
\nonumber\\
&&
=\int_{-\infty}^\infty\intspace dx
\left[
G(x,l)-\left\{1- \exp\left(-e^{-x+Kl}\right) \right\}
\right]
\nonumber\\
&&\sim m(l)-Kl .
\end{eqnarray}
Here, in the last line, we have used the fact that
the wavefront of $G(x,l)$ is located at $x\sim m(l)$,
whereas the wave front of $1- \exp\left(-e^{-x+Kl}\right)$
at $x\sim Kl$.
The distribution function $P(u,l)$ is broad especially in the 
case of a pulled front $(\mu>1/\sqrt{D})$, and the typical 
value corresponds to the maximum of the distribution function.
The typical $u_{typ}$ may define typical values of $y_{q,typ}$
through the relation (\ref{MelTra}),
\begin{eqnarray}
y_{q,typ}\sim e^{[m(l)-Kl]q} .
\label{TypFug}
\end{eqnarray}
Now we can define the dynamical scaling exponents $z_q$
from the $l$-dependence of the typical values $y_{q,typ}(l)$.
To this end, let us recall the following fact.
In a weak disorder regime,
the dynamical exponent of $y_1$ is, as expected, 
just $2-x_1$, where $x_1$ is the scaling dimension of 
$\cos n_a\phi_a$ with $\bfn\in \bfN_1$.
In a strong disorder regime, however, 
it has been shown \cite{CarDouN,CarDouE} that 
the interaction terms in Eq. (\ref{Act}) with the
initial condition $y_n(0)=y_1\delta_{n,1}$
plays a crucial role in the RG flow, and therefore
it is quite important
to keep the operator $y_1\sum\cos n_a\phi_a$ in 
the action and to derive the scaling equation of $y_1$
coupled together with those of higher $y_n$
generated by the fusion of the vertex operators.

Similarly, if one wants to know the dynamical scaling exponent
of the higher $y_q$, 
one has to perturb the action 
by adding the corresponding operator 
$y_q\sum\cos n_a\phi_a$ ($\bfn\in \bfN_q$) as a source term. 
Namely, one has to consider the initial condition
$y_{n}(0)=y_q\delta_{n,q}$,
and hence $\mu=q$ in Eq. (\ref{FroVel}). 
Thus, we find
$y_{q,typ}(l)\sim e^{z_ql}$, where
$z_q$ is defined by
\begin{eqnarray}
z_q=
\left\{
\begin{array}{ll}
2-Kq+g K^2q^2 \quad &\mbox{for}\quad g K^2<2/q^2 ,\\
(\sqrt{8g}-1)Kq \quad &\mbox{for}\quad g K^2\ge2/q^2 .
\end{array}
\right.
\label{DynExp}
\end{eqnarray}
Here we have neglected the logarithmic corrections 
appearing in Eq. (\ref{FroVel}).
The vertex operator 
$\cos(n_a\phi_a)$ with total charge $q$ i.e, $\bfn\in\bfN_q$,
therefore, obeys the scaling law with scaling exponent
$2-z_q$.

In passing, we mention that the logarithmic corrections 
in Eq. (\ref{FroVel})
are universal and give corrections to
the correlation functions like marginal perturbations,
although we have neglected them in this section.
We will briefly discuss them in the next Sec. \ref{s:Sum}, 
calculating logarithmic corrections to the DOS.

\subsection{Calculation of the DOS and the IPR}

By using the dynamical scaling exponents, we first calculate 
the DOS defined in Eq. (\ref{DefDosEtc}).
Although we are basically interested in the Dirac fermion model,
$K=1$, we will derive formulas below using generic $K$,
because they are in fact valid for $K\stackrel{\textstyle <}{\sim}1$. 
Notice that the energy $E$ has the same dimension as $y_1$. 
Then, we find
\begin{eqnarray}
\frac{\Lambda}{E}\sim
\left\{\begin{array}{ll}
e^{(2-K+g K^2)l} & \mbox{for}\quad g K^2<2 ,\\
e^{(\sqrt{8g}-1)Kl} & \mbox{for}\quad g K^2\ge2 ,
\end{array}
\right. 
\label{EL}
\end{eqnarray}
where $\Lambda$ is a renormalized energy, and 
in what follows, we set $\Lambda=1$ for simplicity.
On the other hand, 
since the scaling exponent of $\Gamma^{(1)}$ is $2-z_1$, we have
\begin{eqnarray}
\rho(l)\sim
\left\{\begin{array}{ll}
e^{-(K-g K^2)l} & \mbox{for}\quad g K^2<2 ,\\
e^{-(2+K-K\sqrt{8g})l} & \mbox{for}\quad g K^2\ge2 .
\end{array}
\right. 
\end{eqnarray}
This equation, together with Eq. (\ref{EL}), yields
\begin{eqnarray}
\rho(E)\sim
\left\{\begin{array}{ll}
E^{\frac{K-g K^2}{2-K+g K^2} }
& \mbox{for}\quad g K^2<2 ,\\
E^{\frac{2-\sqrt{8g}K+K}{(\sqrt{8g}-1)K}} 
& \mbox{for}\quad g K^2\ge2 .
\end{array}
\right.
\end{eqnarray}
This result coincides with the one obtained by a variational 
method.\cite{HorDou}
In the weak coupling regime of the Dirac fermion, 
$g<2$ with $K=1$, this DOS is just the same as that obtained 
via usual weak coupling approaches.\cite{LFSG}

\end{multicols}

Next, let us calculate the IPR
according to the definition (\ref{DefDosEtc}).
The IPR are expected to show a power law behavior
$P^{(q)}\sim L^{-\tau(q)}$.
Since the energy $E$ scales as (\ref{EL}), the IPR obey
the scaling law $P^{(q)}(E)\sim E^{\tau(q)/z_q}$
as a function of the energy.
One can read in this way the exponent $\tau(q)$ from $P^{(q)}(E)$
in what follows.
In the case of $g K^2<2$, the $q$th moment scales as
\begin{eqnarray}
\omega^{q-1}\overline{\Gamma^{(q)}}\sim 
\left\{\begin{array}{ll}
e^{-(q-1)(2-K+g K^2)l-(Kq-g K^2q^2)l} 
&
\mbox{for}\quad q<\sqrt{2/g K^2} ,
\\
e^{-(q-1)(2-K+g K^2)l-(2-\sqrt{8g}Kq+Kq)l} 
&
\mbox{for}\quad q\ge\sqrt{2/g K^2 }  ,
\end{array}
\right. 
\end{eqnarray}
where we have used the fact that $\omega$ obeys the same scaling law
as $E$ in Eq. (\ref{EL}).
In the same way as the DOS, we reach
\begin{eqnarray}
P^{(q)}(E)\sim
\left\{\begin{array}{ll}
E^{\frac{(q-1)(2-g K^2q)}{2-K+g K^2} }
& \mbox{for}\quad q<\sqrt{2/g K^2} ,\\
E^{\frac{2q(1-\sqrt{g/2}K)^2}{2-K+g K^2}} 
& \mbox{for}\quad q\ge\sqrt{2/g K^2} .\\
\end{array}
\right. 
\quad  (g K^2<2).
\label{Pq1}
\end{eqnarray}
Contrary to this, in the case $g K^2\ge2$, we have
\begin{eqnarray}
\omega^{q-1}\overline{\Gamma^{(q)}}\sim 
\left\{\begin{array}{ll}
e^{-(q-1)(\sqrt{8g}-1)Kl-(Kq-g K^2q^2)l} 
&
\mbox{for}\quad q<\sqrt{2/g K^2} ,
\\
e^{-(q-1)(\sqrt{8g}-1)Kl-(2-\sqrt{8g}Kq+Kq)l} 
&
\mbox{for}\quad q\ge\sqrt{2/g K^2} .\\
\end{array}
\right. 
\end{eqnarray}
Therefore, we arrive at
\begin{eqnarray}
P^{(q)}(E)\sim
\left\{\begin{array}{ll}
E^{-\frac{2(1-\sqrt{g/2}Kq)^2}{(\sqrt{8g}-1)K} }
& \mbox{for}\quad q<\sqrt{2/g K^2} ,\\
E^0 
& \mbox{for}\quad q\ge\sqrt{2/g K^2} .\\
\end{array}
\right. 
\quad (g K^2\ge2).
\label{Pq2}
\end{eqnarray}
These results completely reproduce those obtained so far
by using various methods.\cite{CMW,CCFGM,CarDouE}

The merit of the present approach lies in the fact that 
spatial correlations are quite easy to calculate.
We assume that $q_1\ge q_2$ without loss of generality.
Noticing Eq. (\ref{GreBos}) leads to
\begin{eqnarray}
Q^{(q_1,q_2)}(x-y,E)&&\sim
\omega^{q_1+q_2-1}\overline{\Gamma^{(q_1)}(x)\Gamma^{(q_2)}(y)}
/\rho(E)
\nonumber\\
&&\sim
\frac{1}{ |x-y|^{\Delta(q_1,q_2)} }
\omega^{q_1+q_2-1}\overline{\Gamma^{(q_1+q_2)}(y)}/\rho(E) 
\nonumber\\
&&
\sim
\frac{1}{ |x-y|^{\Delta(q_1,q_2)} }P^{(q_1+q_2)}(E) ,
\label{Qq1q2}
\end{eqnarray}
where we have kept the most relevant operator for the OPE in the 
second line.
Some comments are in order:
First, the OPE used here is valid in the energy scale
$|x-y|\ll E^{-z_1}$, where $z_1$ is the dynamical exponent in 
Eq. (\ref{DynExp}).
Next, the formula (\ref{Pq1}) or (\ref{Pq2}) applies to
$P^{(q_1+q_2)}(E)$ in the last line of (\ref{Qq1q2}).
Finally, the exponent $\Delta(q_1,q_2)$ is calculated as
\begin{eqnarray}
\Delta(q_1,q_2)&&=(2-z_{q_1})+(2-z_{q_2})-(2-z_{q_1+q_2})
\nonumber\\
&&=
\left\{
\begin{array}{ll}
2g K^2q_1q_2  
&
\mbox{for}\quad g K^2\le2/(q_1+q_2)^2 ,
\\
\sqrt{8g}K(q_1+q_2)-g K^2(q_1^2+q_2^2)-2 
&
\mbox{for}\quad 2/(q_1+q_2)^2\le g K^2\le2/q_1^2 ,
\\
\sqrt{8g}Kq_2-g K^2q_2^2 
&
\mbox{for}\quad 2/q_1^2\le g K^2\le2/q_2^2 ,
\\
2 
&
\mbox{for}\quad 2/q_2^2\le g K^2 .
\end{array}
\right.
\label{SpaCorExp}
\end{eqnarray}
The equation shows that if disorder is strong enough, 
the exponent $\Delta(q_1,q_2)$
saturates at 2 for any $q_1$ and $q_2$.
This saturation implies that the model is in the fully frozen phase.

\begin{multicols}{2}

Ryu and Hatsugai\cite{RyuHat} have recently calculated the density 
correlation function of the zero energy state of the Dirac fermion model, 
which corresponds to $q_1=q_2=1$ and $K=1$ case.
Substituting these values into Eq. (\ref{SpaCorExp}), we have
\begin{eqnarray}
\Delta=
\left\{
\begin{array}{ll}
2g 
&
\mbox{for}\quad g \le 1/2 ,
\\
2\left(\sqrt{8g}-g-1\right) 
&
\mbox{for}\quad 1/2\le g \le2 ,
\\
2 
&
\mbox{for}\quad 2\le g .
\end{array}
\right.
\end{eqnarray}
It should be noted that if one compares this formula directly 
with the calculation of Ryu and Hatsugai, one has to replace 
$g\rightarrow g/\pi$.
The above formula seems to fit their calculation, and therefore
we can claim that an evidence of the strong disorder regime 
has actually revealed itself in their numerical calculation
of the exponent.

\section{Summary and discussions}\label{s:Sum}

In this paper, we have studied the Dirac fermion model with
the random vector field in order to calculate the DOS, the IPR, and
their spatial correlations both in the strong
and the weak disorder regime.
We have derived scaling equations of one-loop order, 
using the OPE techniques as well as
taking account of the fusion of the vertex operators.
We have been able to reproduce the results known so far.
Especially, we have shown that  
the numerical calculation of the density
correlation function
can be explained by the RG method in this paper.
It turns out that the saturation of the exponent at 2 
is manifestly due to the freezing transition.

So far we have studied the random Dirac fermion in the continuum limit.
In this case, 
the fermion has of course chiral symmetry at any realization 
of disorder.
On the other hand, starting from the lattice model, e.g, 
the random hopping fermion on a square lattice with $\pi$-flux 
per plaquette,\cite{Hat}
or on a honeycomb lattice,\cite{Sem} 
we find, in a continuum limit, 
a Dirac fermion with two flavors due to the species doubling.
These lattice models could be described by such a Dirac 
fermion with a random imaginary vector field 
as well as a random mass term,\cite{GLL}
which has indeed chiral symmetry.
The pioneering work by Gade\cite{Gad} of the disorder systems with
chiral symmetry has predicted that
the DOS diverges toward the zero energy
as $E^{-1}\exp(-c|\ln E|^x)$ with $x=1/2$.
Recently, Motrunich, Damle, and Huse\cite{MDH} 
have pointed out that the DOS of such models
should obey the the same scaling law but with $x=2/3$
due to strong disorder effects.
Surprisingly, Mudry, Ryu and Furusaki\cite{MRF} have 
quite recently shown that 
such a behavior is also due to the freezing transition.
Their method is based on the RG equation of the DOS,
by taking account of all relevant perturbations. 
The KPP equation also plays a crucial role in their analysis.

In this sense, the KPP equation, which was proposed originally as
a biological problem of the gene,\cite{KPP}
is a key equation to understand various kinds
of random systems such as spin glasses,\cite{Der}
a directed polymers on a Cayley tree,\cite{DerSpo}
and a random hopping fermion on a lattice.\cite{MRF}

Although we claimed in Sec. \ref{s:DOS} that the logarithmic 
corrections in Eq. (\ref{FroVel}) are universal, 
we have neglected them for simplicity.
It is, however, readily seen that they give the logarithmic corrections,
for example, to the DOS as \cite{CarDouE}
\begin{eqnarray}
\rho(E)\sim
E^{ \frac{2-(\sqrt{8g}-1)K}{(\sqrt{8g}-1)K} }
\left|\ln E\right|^{-2\alpha\frac{\sqrt{8g}}{\sqrt{8g}-1}} ,
\end{eqnarray}
where 
\begin{eqnarray}
\alpha=
\left\{
\begin{array}{ll}
\frac{1}{8}
&
\mbox{for}\quad gK^2=2 ,
\\
\frac{3}{8} \quad
&
\mbox{for}\quad gK^2>2 .
\end{array}
\right. 
\end{eqnarray}
In deriving the equation, we have included the logarithmic 
corrections into the dynamical scaling exponents
via the relation $z_q=(\partial_lu_{typ})q=[\partial_lm(l)-K]q$.
In the weak disorder regime, 
however, no logarithmic corrections should exist
in the DOS, as can be seen from Eq. (\ref{FroVel}).

In pure systems without disorder, it is well-known that
marginal perturbations give rise to such logarithmic corrections.
In random systems, however, logarithmic corrections are allowed 
even in the fixed-point theories themselves, due to
the presence of so-called logarithmic operators.\cite{Gur}
In the present case, it may be quite interesting to address the 
question on the origin of the logarithmic corrections
appearing in the KPP equation.

\acknowledgements
The author would like to thank C. Mudry for fruitful discussions
and a plenty of comments.
He is supported in part by the Japan Society for 
the Promotion of Science and the Yamada Science Foundation.


\end{multicols}

\end{document}